\documentclass[12pt]{article}

\usepackage{amssymb}
\usepackage{amsmath}
\usepackage{amscd}
\usepackage{latexsym}
\usepackage{graphicx}
\usepackage{url}

\usepackage{cite}

\newcommand{\be}{\begin{equation}}
\newcommand{\ee}{\end{equation}}

\newcommand{\dlt}{\delta}

\newcommand{\bt}{\beta}

\newcommand{\al}{\alpha}
\newcommand{\ra}{\rightarrow}
\newcommand{\sgm}{\sigma}

\newcommand{\Gm}{\Gamma}

\newcommand{\cB}{{\cal B}}
\newcommand{\cH}{{\cal H}}
\newcommand{\cL}{{\cal L}}
\newcommand{\cA}{{\cal A}}

\newcommand{\rgl}{\rangle}
\newcommand{\lgl}{\langle}

\begin{document}

\begin{center}

{\Large{\bf Quantum probability and quantum decision making} \\ [5mm]

V.I. Yukalov$^{1,2,*}$ and D. Sornette$^{1,3}$} \\ [3mm]

{\it
$^1$Department of Management, Technology and Economics, \\
ETH Z\"urich (Swiss Federal Institute of Technology) \\
Scheuchzerstrasse 7,  Z\"urich CH-8032, Switzerland \\ [3mm]

$^2$Bogolubov Laboratory of Theoretical Physics, \\
Joint Institute for Nuclear Research, Dubna 141980, Russia \\ [3mm]

$^3$Swiss Finance Institute, c/o University of Geneva, \\
40 blvd. Du Pont d'Arve, CH 1211 Geneva 4, Switzerland}
\end{center}

\vskip 3cm

\begin{abstract}
\begin{sloppypar}
A rigorous general definition of quantum probability is given, which is 
valid for elementary events and for composite events, for operationally 
testable measurements as well as for inconclusive measurements, 
and also for non-commuting observables in addition to commutative 
observables. Our proposed definition of quantum probability makes it possible 
to describe quantum measurements and quantum decision making on the same 
common mathematical footing. Conditions are formulated for the case when 
quantum decision theory reduces to its classical counterpart and for the 
situation where the use of quantum decision theory is necessary. 
\end{sloppypar}
\end{abstract}

{\parindent=0pt
\vskip 2cm
{\bf Keywords}: quantum probability, quantum measurements, quantum decision making

\vskip 5cm
{\bf Authors for correspondence}: V. I. Yukalov\\
{\bf E-mail}: yukalov@theor.jinr.ru 
}

\newpage

\section{Introduction}

A general and mathematically correct definition of quantum probability is 
necessary for several important applications: theory of quantum measurements, 
theory of quantum information processing and quantum computing, quantum 
decision theory, and creation of artificial quantum intelligence. Although the
definition of quantum probability for operationally testable events is well
known and used from the beginning of quantum theory \cite{Neumann_1}, such a 
definition for composite events, corresponding to noncommuting observables, has
been a long-standing problem. This problem becomes especially important in the
application of the quantum approach to psychological and cognitive sciences, 
where there exist not only operationally testable events, but also decisions 
under uncertainty, corresponding to operationally uncertain events. Moreover, 
for decision making in real-life, decisions under uncertainty are not exceptions, 
but rather are common typical situations.  

Classical decision theory, based on the notion of utility \cite{Neumann_2},
is known to yield numerous paradoxes in realistic decision making 
\cite{Machina_3}. This is why a variety of quantum models has been suggested
for applications in psychological and cognitive sciences, as can be inferred
from books \cite{Khrennikov_4,Busemeyer_5,Bagarello_6,Haven_7} and reviews 
\cite{YS_8,Sornette_9,Ashtiani_10}. 

Applying quantum theory to psychological and cognitive sciences, researchers 
have often constructed special models designed specifically to treat particular 
cases of decision making. However, to our firm understanding, the theory of 
quantum decision making has to be formulated as a general theory valid for 
arbitrary cases. Moreover, such a theory should have the same mathematical 
grounds as the theory of quantum measurements. Really, the latter can be 
interpreted as decision theory \cite{Neumann_1}. Between measurements and 
decisions, there is a direct correspondence requiring just a slight language 
change: measurements correspond to events; operationally testable measurements 
are analogous to certain events; undefined measurements can be matched to 
uncertain events; composite measurements are equivalent to composite decisions. 

The aim of this paper is to present a general theory, with a unique well-defined
mathematical basis, which would be valid for both quantum measurements as 
well as for quantum decision making. The main point of such an approach lies in
a correct definition of quantum probability that would be applicable for any 
type of measurements and events, operationally testable or inconclusive, 
elementary or composite, corresponding to commuting or noncommuting 
observables. The theory has to be valid for closed as well as for open systems, 
for individual as well as for social decision makers. Also, it has to be more 
general than classical theory, including the latter as a particular case and 
clearly distinguishing the conditions necessarily requiring the use of quantum 
techniques and those when the classical approach is sufficient. Finally, it should 
not be just a descriptive way of modeling, but it must allow for quantitative 
predictions.

\section{Main preliminary notions}

\subsection{Quantum-classical correspondence principle}

In order to constrain and anchor the general quantum theory, we require the 
validity of the {\it quantum-classical correspondence principle}. This 
principle was put forward by Bohr \cite{Bohr_11,Bohr_46} for a particular 
case related to atomic spectra. Later its applicability was extended to other 
problems of quantum mechanics, with the Ehrenfest equations being one of the 
illustrations \cite{Schiff_47}. Nowadays, this principle is understood in the 
generalized sense as the requirement that classical theory be a particular 
case of quantum theory \cite{Zurek_48}. In the present context, it implies 
that the theory of quantum measurements should include the theory of classical 
measurements, that quantum decision theory should include classical decision 
theory, and that classical probability should be a particular case of quantum 
probability.      

In what follows, we use the term {\it event}, implying that this can be an event
in decision theory or probability theory, or the result of a measurement in 
the quantum theory of measurements.

\subsection{Quantum logic of events}

The algebra of events is prescribed by quantum logic \cite{Birkhoff_12}. 
Events form an event ring $\mathcal{R} = \{ A_i: i = 1,2,\ldots\}$ possessing 
two binary operations, addition and conjunction. Addition is 
such that for any $A, B \in \mathcal{R}$, there exists 
$A \cup B \in \mathcal{R}$ with the properties:
$$
A \cup B = B \cup  A  \qquad ( commutativity) \;  ,
$$
$$
 A \cup \left ( B  \cup C \right ) = \left ( A  \cup B \right )
\cup C \qquad (associativity) \; ,
$$
$$
A \cup A = A \qquad (idempotency) \;   .
$$
Conjunction means that for any $A, B \in \mathcal{R}$, there exists 
$A \cap B \in \mathcal{R}$ satisfying the properties:
$$
\left ( A  \cap B \right )
\cap C = A \cap  \left ( B  \cap C \right ) 
\qquad (associativity) \; 
$$
$$
 A \cap A = A \qquad (idempotency) \;  .
$$
But, generally, conjunction is not commutative and not distributive:
$$
A \cap B \neq B \cap  A  \qquad (no \; commutativity) \; ,
$$
$$
 A \cap \left ( B  \cup C \right ) \neq \left ( A  \cap B \right )
\cup A \cap C \qquad (no \; distributivity) \; .
$$

The fact that distributivity is absent in quantum logic was emphasized by
Birkhoff and von Neumann \cite{Birkhoff_12}, who illustrated this by the 
following example. Suppose there are two events $B_1$ and $B_2$ that, when 
combined, form unity, $B_1 \cup B_2 = 1$. Moreover, $B_1$ and $B_2$ are 
such that each of them is orthogonal to a nontrivial event $A \neq 0$, hence 
$A \cap B_1 = A \cap B_2 = 0$. According to this definition, 
$A \cap (B_1 \cup B_2) = A \cap 1 = A$. But if the property of 
distributivity were true, then one would get 
$(A \cap B_1) \cup (A \cap B_2) = 0$. This implies that $A = 0$,  
which contradicts the assumption that $A \neq 0$. 

It is easy to illustrate the concept of non-distributivity in quantum 
physics by numerous examples. The simplest of these is as follows 
\cite{Hughes_13}. Let us measure the spin projection of a particle with 
spin $1/2$. Let $B_1$ be the event of measuring the spin in the up state with 
respect to the axis $z$, while $B_2$ is the event of measuring the spin in 
the down state along this axis. Since the spin can be either up or down, 
$B_1 \cup B_2 = 1$. And let $A$ be the event of measuring the spin along 
an axis in the plane orthogonal to the axis $z$. According to the rules of 
quantum mechanics, the spin cannot be measured simultaneously along two 
orthogonal axes, it is found either measured along one axis or along another 
axis but cannot have components on both axes at the same time. Hence 
$A \cap B_1 = A \cap B_2 = 0$, while $A \cap (B_1 \cup B_2) \neq 0$.
Therefore, there is no distributivity of events in the spin measurement. 

Thus the non-distributivity of events is an important concept that should not 
be forgotten in applying quantum theory to cognitive sciences.

\subsection{Decision maker state}

In quantum theory, systems can be closed or open. Respectively, their states 
can be described by wave functions or as statistical operators. How should
one interpret the state of a decision maker, as a wave function or as a
statistical operator? Such a state, characterizing the given decision maker,
can be called a {\it strategic decision-maker state} 
\cite{YS_14,YS_15,YS_16}. 

Recall the notion of an isolated system in quantum theory. Strictly speaking,
quantum systems cannot be absolutely isolated, but can only be 
{\it quasi-isolated} \cite{Yukalov_17,Yukalov_18}, which means the following.  
At initial time $t = 0$, one can prepare a system in a pure state described 
by a wave function. However, there always exist uncontrollable external 
perturbations or noise from the surrounding, resulting in the system decoherence 
beyond a time $t_{dec}$, which makes the system state mixed. Also, to confirm 
that the considered system is to some extent isolated, it is necessary to check 
this by additional control measurements starting at time $t_{con}$, which 
again disturbs the system's isolation. In this way, one can assume that the 
system is quasi-isolated during the interval of time 
$0 < t < min \{t_{dec}, t_{con}\}$. 

Decision makers, generally, are the members of a society, hence, they 
correspond to non-isolated open systems that have to be described by 
statistical operators. One could think that in laboratory tests, it would 
be admissible to treat decision makers as closed systems and to characterize 
them by wave functions. This, however, is not correct. First of all, in 
laboratory tests, even when being separated from each other, decision makers 
do communicate with the investigators performing the test. Moreover, even when 
being for some time locked in a separate room, any decision maker possesses 
the memory of interacting with many people before as well as his/her 
expectations of future interactions, which influences his/her decisions. From 
the physiological point of view, {\it memory is nothing but delayed interactions}. 
Therefore, no decision maker can be treated as an isolated system, which 
excludes the validity of using a wave function description. The correct 
treatment of any decision maker requires to consider him/her as an open 
system, hence, characterized by a statistical operator.

\subsection{Operationally testable events}

In the theory of quantum measurements or quantum decision theory, the simplest 
case occurs when one deals with a simple event corresponding to a single 
measurement, or a single action. Observable quantities in quantum theory are 
represented by self-adjoint operators, say $\hat{A}$, from the algebra of 
local observables. Measuring an eigenvalue $A_n$ of the operator can be 
interpreted as the occurrence of an event $A_n$. The corresponding eigenvector 
$|n\rangle$ is termed a microstate in physics, or event mode in decision theory. 
Here and in what follows, the family of eigenvectors is assumed to be 
orthonormalized. Respectively, the operator 
$\hat{P}_n \equiv |n \rangle \langle n|$ is a measurement projector in physics, 
or an event operator in decision theory. The collection $\{\hat{P}_n\}$ is 
a projector-valued measure.   
   
The space of microstates, or the space of decision modes, is given by the
Hilbert space 
\be
\label{1}
  \cH_A = {\rm span} \{ | n \rgl \} \; .
\ee
The considered quantum system state, or decision maker strategic state, 
is characterized by a statistical operator $\hat{\rho}$. The pair 
$\{\mathcal{H},\hat{\rho}\}$ is a statistical ensemble, or decision 
ensemble. The probability of measuring an eigenvalue $A_n$, or the 
probability of an event $A_n$, is given by the formula
\be
\label{2}
 p(A_n) = {\rm Tr}_A \hat \rho \hat P_n \equiv 
\lgl \hat P_n \rgl \;  ,
\ee
where the trace operation is over space (\ref{1}). This probability is 
uniquely defined for any Hilbert space (\ref{1}) of dimensionality larger 
than two \cite{Gleason_19}.

\subsection{Problem of degenerate spectrum}

The spectrum of the considered operator can happen to be degenerate, which 
implies that a single eigenvalue $A_n$ corresponds to several eigenvectors 
$|n_j\rangle$, with $j = 1,2,\ldots$. Does this create any problem?

This is not a problem in quantum measurements. In the case of degeneracy,
one introduces a projector
\be
\label{3}
\hat P_n \equiv \sum_j \hat P_{n_j} \qquad 
\left ( \hat P_{n_j} = | n_j \rgl \lgl n_j | \right ) \; ,
\ee
so that the probability of measuring $A_n$ becomes
\be
\label{4}
p(A_n) = {\rm Tr}_A \hat\rho \hat P_n = 
\sum_j \lgl \hat P_{n_j} \rgl \;   .
\ee
  
Degeneracy may seem to be an annoyance in decision theory. Really, if $A_n$ 
is a degenerate event related to a degenerate spectrum, then what would be
the meaning of the different modes associated with the same event? It is 
necessary to ascribe some meaning to these different modes, otherwise the 
situation will be ambiguous.  

Fortunately, the problem of degeneracy is easily avoidable, both in physics
as well as in decision theory. In physics, degeneracy can be lifted by 
switching on arbitrarily weak external fields. In decision theory, this 
would correspond to reclassifying the events by adding small differences 
between the events. Mathematically, the procedure of lifting degeneracy is 
done by adding to the considered operator of an observable an infinitesimally
small term breaking the symmetry that caused the degeneracy, which means the
replacement
\be
\label{5}
 \hat A \rightarrow \hat A + \nu \hat\Gm \qquad (\nu \ra 0) \;  .
\ee
The related eigenvalues $A_{n_j} + \nu \Gamma_{n_j}$ become nondegenerate. 
Then the probability of each subevent can be defined as
\be
\label{6}
 p(A_{n_j} ) = \lim_{\nu\ra 0} p\left ( A_{n_j} + 
\nu \Gm_{n_j} \right ) \;  .
\ee
Such a procedure of degeneracy lifting was mentioned by von Neumann 
\cite{Neumann_1} for quantum systems and developed as the method of 
quasi-averages by Bogolubov \cite{Bogolubov_20,Bogolubov_21} for statistical 
systems.    

In any case, neither in physics nor in decision theory, the problem of 
spectrum degeneracy is actually a principal problem. One just needs to 
either ascribe a meaning to different modes of an event, or one can avoid 
the problem completely by lifting the degeneracy, which corresponds to a 
reclassification of events, as already mentioned. The latter way is 
preferable in decision theory, since it avoids the ambiguity in dealing 
with unspecified degeneracy.

\subsection{Consecutive quantum measurements}

In quantum theory, one considers the possibility of measuring two 
observables immediately one after the other. The standard treatment of this
process is as follows. Suppose, first, one accomplishes a measurement for
an observable represented by an operator $\hat{B}$, with eigenvalues 
$B_\alpha$ and eigenvectors $| \alpha \rangle$. The event $B_\alpha$ is 
represented by the projector 
$\hat{P}_\alpha \equiv |\alpha \rangle \langle \alpha|$. One assumes that, 
immediately after measuring $B_\alpha$, the system state reduces from 
$\hat{\rho}$ to the state
\be
\label{7} 
\hat\rho_\al \equiv 
\frac{\hat P_\al \hat\rho \hat P_\al}{{\rm Tr}\hat\rho\hat P_\al} \; .
\ee

Immediately after the first measurement, one accomplishes a measurement for
an observable represented by an operator $\hat{A}$, with eigenvalues $A_n$
and eigenvectors $|n \rangle$. The event $A_n$ is represented by the projector
$\hat{P}_n \equiv |n \rangle \langle n|$. 

The probability of these consecutive measurements is the L\"{u}ders 
\cite{Luders_22} probability
\be
\label{8}
 p_L(A_n | B_\al ) \equiv {\rm Tr}\hat\rho_\al \hat P_n =
 \frac{{\rm Tr}\hat\rho\hat P_\al \hat P_n \hat P_\al}{{\rm Tr}\hat\rho\hat P_\al} \; ,
\ee
also called von Neumann-L\"{u}ders probability. 

By introducing the Wigner \cite{Wigner_23} probability
\be
\label{9}
 p_W(A_n | B_\al ) \equiv 
{\rm Tr} \hat\rho \hat P_\al \hat P_n \hat P_\al \;  ,
\ee
one comes to the relation
\be
\label{10}
 p_W(A_n | B_\al ) = p_L(A_n | B_\al ) p(B_\al) \; .
\ee
This formula is reminiscent of the relation between the joint probability 
of two events and the conditional probability for these events. Because 
of this similarity, one interprets the Wigner probability $p_W$ as a joint 
probability and the L\"{u}ders probability $p_L$ as a conditional 
probability.  

However, by direct calculations, assuming nondegenerate events, we have
\be
\label{11}
 p_L(A_n | B_\al )  = | \lgl n | \al \rgl |^2 \;  .
\ee
This form is symmetric with respect to the interchange of 
events. Therefore the L\"{u}ders probability cannot be treated as 
the generalization of the classical conditional probability that is not 
necessarily symmetric. Respectively, the Wigner probability cannot be 
considered as a joint probability of two events \cite{YS_24}. 

One could think that, by invoking degenerate events, it would be possible
to avoid the problem. Suppose the events $A_n$ and $B_\alpha$ are 
degenerate, so that their projectors are
\be
\label{12}
 \hat P_n = \sum_i \hat P_{n_i} \; , \qquad
 \hat P_\al = \sum_j \hat P_{\al_j} \;  .
\ee
Then we have
\be
\label{13}
 p_W(A_n | B_\al ) = \sum_{ijk} \lgl \al_i | \hat \rho | \al_j \rgl
 \lgl \al_j | n_k \rgl  \lgl n_k | \al_i \rgl \; , \qquad
p(B_\al) = \sum_j \lgl \al_j | \hat \rho | \al_j \rgl \; .
\ee
Interchanging the events yields
\be
\label{14}
 p_W(B_\al | A_n ) = \sum_{ijk} \lgl n_i | \hat \rho | n_j \rgl
 \lgl  n_j | \al_k \rgl  \lgl \al_k | n_i \rgl \; , \qquad
p(A_n) = \sum_j \lgl n_j | \hat \rho | n_j \rgl \;  .
\ee
This shows that the L\"{u}ders probability, generally, is not symmetric
for degenerate events. 

But let us remember the quantum-classical correspondence principle, according
to which classical theory has to be a particular case of quantum theory. In 
classical theory, the field of events is commutative. In quantum theory, 
commuting observables share the same family of eigenvectors. This can be 
formulated as the property $\lgl \al_i | n_j \rgl = \dlt_{ij} \dlt_{\al\bt}$.
Then, passing to commutative events, for the L\"{u}ders probability (\ref{8}) 
we obtain
\be
\label{15}
 p_L(A_n | B_\al ) = \dlt_{n\al} = p_L(B_\al| A_n ) \; .
\ee
This is not merely symmetric, but even trivial. Contrary to this, classical
conditional probabilities are neither symmetric nor trivial. 

Thus, the quantum-classical correspondence principle does not hold, which means 
that the L\"{u}ders probability in no way should be accepted as a generalization
of classical conditional probability. The L\"{u}ders probability is just a 
transition probability. If one wishes, one can use it as a transition probability 
in the frame of a narrow class of physical measurements. However, it is not a 
conditional probability in the general sense, and its use as such for cognitive 
sciences is not correct \cite{YS_24,YS_25,Boyer_26}. 

It is worth mentioning that the Kirkwood \cite{Kirkwood_27} form
$\lgl\hat P_n\hat P_\al\rgl={\rm Tr} \hat \rho \hat P_n \hat P_\al$
also cannot be accepted as a probability, since it is complex-valued.    

Concluding this section, we stress that the standard von Neumann-L\"{u}ders 
transition probability cannot be treated as a generalization of classical 
conditional probability to the quantum region, since it does not satisfy the
quantum-classical correspondence principle. And the consideration of 
degenerate events does not save the situation.

\subsection{Realistic measurement procedure}

The problem with the von Neumann-L\"{u}ders probability lies in its 
oversimplified nature, giving only a cartoon of the much more complicated
procedure of realistic measurements. This cartoon ignores the existence and 
influence of a measuring device, it ignores the finite time of any 
measurement, and it ignores that during measurements and between them, 
the system evolves. The correct description of a realistic measurement 
procedure is as follows \cite{YS_24}. 

Let us assume that we are interested in measuring two observables 
corresponding to the operators $\hat{A}$ and $\hat{B}$, with eigenvalues 
$A_n$ and $B_\alpha$ and eigenvectors $|n \rangle$ and $|\alpha \rangle$,
respectively. The related event representations are
\be
\label{16}
 A_n \ra | n \rgl \ra \hat P_n = | n \rgl \lgl n | \; , \qquad
 B_\al \ra | \al \rgl \ra \hat P_\al = | \al \rgl \lgl \al | \;  .
\ee
According to Eq. (\ref{1}), the corresponding mode spaces are
\be
\label{17}
\cH_A \equiv {\rm span}\{ | n \rgl \} \; , \qquad
 \cH_B \equiv {\rm span}\{ | \al \rgl \} \;  .
\ee
  
To measure anything, one needs a measuring device, whose internal states
are the vectors of a Hilbert space $\mathcal{H}_M$. In decision theory, this
state corresponds to internal states of a decision maker. The total space,
containing all possible microstates, is the tensor-product space
\be
\label{18}
\cH = \cH_A \bigotimes \cH_B \bigotimes \cH_M \;   .
\ee

The measurement procedure consists of several channels. The first step of 
any measurement is the preparation of the device for measurement, which 
can be represented by the entangling channel
\be
\label{19}
C_1 : \; \hat\rho_A(0) \bigotimes \hat\rho_B(0) \bigotimes 
\hat\rho_M(0) \ra \hat\rho(t_1) \;   ,
\ee
describing the formation from initial partial states, during the preparation 
time $t_1$, of an entangled total state $\hat{\varrho}(t_1)$ of the system 
plus the measuring device.  
 
Before the measurement starts, the total state evolves until time $t_2$,
according to the channel
\be
\label{20}
 C_2 : \; \hat\rho(t_1) \ra \hat\rho(t_2) \; ,
\ee
where
$$
 \hat\rho(t_2) = 
\hat U(t_2 - t_1) \hat\rho(t_1) \hat U^+(t_2 - t_1) \; ,
$$
with $\hat{U}$ being the evolution operator. 

In the interval of time $[t_2,t_3]$, one measures the observable 
corresponding to the operator $\hat{B}$, which is described by the 
partially disentangling channel
\be
\label{21}
 C_3 : \; \hat\rho(t_2) \ra 
\hat\rho_{AM}(t_3) \bigotimes \hat\rho_B(t_3) \;  ,
\ee
where
$$
 \hat\rho_{AM}(t_3) = {\rm Tr}_B \hat\rho(t_3) \; , \qquad
\hat\rho_{B}(t_3) = {\rm Tr}_{AM} \hat\rho(t_3) \;   .
$$
Disentangling, or separating $\hat\rho_{B}(t_3)$ from the total state 
is necessary for measuring the values related to the operator of the 
observable $\hat{B}$. According to the standard definition, separating 
a subsystem implies tracing out all other degrees of freedom, except those 
of the considered subsystem. 

Then, until time $t_4$, the system again is getting entangled by the
evolution channel
\be
\label{22}
 C_4 : \; \hat\rho_{AM}(t_3) \bigotimes \hat\rho_B(t_3) \ra
\hat\rho(t_4) \; ,
\ee
where
$$
\hat\rho(t_4) = \hat U(t_4 - t_3) \hat\rho_{AM}(t_3) \bigotimes 
\hat\rho_B(t_3) \hat U^+(t_4 - t_3) \;   .
$$

Finally, in the interval of time $[t_4,t_5]$, one accomplishes a measurement
of the observable associated with the operator $\hat{A}$, which is 
characterized by the partially disentangling channel
\be
\label{23}
 C_5 : \;  \hat\rho(t_4) \ra  \hat\rho_A(t_5) \bigotimes
\hat\rho_{BM}(t_5) \; , 
\ee
where
$$
 \hat\rho_A(t_5) = {\rm Tr}_{BM} \hat\rho(t_5)\; , \qquad 
\hat\rho_{BM}(t_5) = {\rm Tr}_A \hat\rho(t_5) \; .
$$

Summarizing, the process of measurement of two observables is a procedure 
represented by the channel convolution
\be
\label{24}
  C = C_5 \bigotimes C_4 \bigotimes C_3 \bigotimes C_2 \bigotimes C_1  
\ee
and consisting of five steps:
\begin{eqnarray}
\label{25}
\begin{array}{cll}
C_1 : ~ & ~ preparation , ~ & ~ t \in [0,t_1] \; , \\
C_2 : ~ & ~ evolution ,   ~ & ~ t \in [t_1,t_2] \; , \\
C_3 : ~ & ~ B - measurement , ~ & ~ t \in [t_2,t_3] \; , \\
C_4 : ~ & ~ evolution , ~   & ~ t \in [t_3,t_4] \; , \\
C_5 : ~ & ~ A - measurement , ~ & ~ t \in [t_4,t_5] \;  .
\end{array}
\end{eqnarray}
The evolution channels are unitary but entangling, while the measurement 
channels are disentangling but nonunitary. The measurement channels are 
nonunitary because they involve the trace operation that cannot be 
represented by a unitary operator. The realistic measurement procedure 
is more complicated than the von Neumann-L\"{u}ders scheme and, generally, 
cannot be reduced to the latter even if the involved intervals of time 
are rather short.

\section{Joint quantum probability}

\subsection{Channel-state duality}

As is explained in Sec. 2.f, the von Neumann-L\"{u}ders scheme does not
provide a general definition of conditional quantum probabilities and 
therefore does not lead to correct joint quantum probabilities. This is 
due to the fact that a realistic measurement procedure requires the 
five-step convolution channels described in the previous section. This 
multichannel measurement procedure looks quite complicated. Fortunately, 
there exists the Choi-Jamiolkowski \cite{Choi_28,Jamiolkowski_29} 
isomorphism establishing the channel-state duality
\be
\label{26}
 C \longleftrightarrow \{ \hat\rho_{AB} , \; \cH_{AB} \} \;  ,
\ee
with a state $\hat\rho_{AB}$ defined on the Hilbert space
\be
\label{27}
 \cH_{AB} = \cH_{A} \bigotimes \cH_{B} \;  .
\ee
Thus, instead of dealing with the channel convolution, we can equivalently
consider the composite state characterized by the space of microstates
(\ref{27}).

\subsection{Prospects as composite events}

Using the channel-state duality, we can interpret the measurement of 
two observables, or the occurrence of two events, as a composite event.
For instance, let us consider events $A$ and $B$. The corresponding
composite event, called {\it prospect}, is $A \bigotimes B$, which is 
represented by the tensor product of two event operators as
\be
\label{28}   
 A \bigotimes B \ra \hat P_A \bigotimes \hat P_B \;  ,
\ee
with the event operators
$\hat P_A = |A \rgl \lgl A|, \; \hat P_B = |B \rgl \lgl B|$. 

The joint probability of the prospect composed of two events is
\be
\label{29}
 p \left (A \bigotimes B \right ) = {\rm Tr}_{AB} \hat\rho_{AB} 
\hat P_A \bigotimes \hat P_B \; .
\ee
This definition has been employed from the beginning of the development
of our approach named Quantum Decision Theory (QDT)
\cite{YS_8,YS_14,YS_15,YS_16,YS_30,YS_31}. We use the term {\it prospect}
for a composite event, since when applying the QDT to decision making,
we calculate the classical part of the quantum probability by invoking
the notion of utility \cite{YS_30,YS_31,YS_49}.

\subsection{Conditional quantum probabilities}

Having defined the joint probability of events, it is straightforward to
introduce the conditional probabilities
\be
\label{30} 
 p( A| B ) \equiv \frac{p(A\bigotimes B)}{p(B)} \; , \qquad
 p( B| A ) \equiv \frac{p(B\bigotimes A)}{p(A)} \; ,
\ee
with the marginal probabilities
\be
\label{31}
p(A) = {\rm Tr}_{AB} \hat\rho_{AB} \hat P_A \bigotimes \hat 1_B \; , \qquad
p(B) = {\rm Tr}_{AB} \hat\rho_{AB} \hat 1_A \bigotimes \hat P_B \;  .
\ee
Here $\hat{1}_A$ and $\hat{1}_B$ are unity operators in the corresponding 
spaces. Clearly, the conditional probabilities, in general, are not symmetric. 

Note that this definition of conditional probabilities is self-consistent
and does not meet the problem of connecting conditional and joint 
probabilities, as in the case when conditional probabilities are defined
through the L\"{u}ders form \cite{Asano_32}.

\subsection{Separable and entangled prospects}

The property of entanglement is important for both quantum measurements 
as well as for quantum decision making \cite{YS_33,YYS_34}.
There are two types of prospects that qualitatively differ from each other,
separable and entangled, whose rigorous definition is given below. 

Let $\cA = \{\hat A\}$ be an algebra of local observables defined
on a Hilbert space $\cH_A$. For any two operators $\hat A_1$ and 
$\hat A_2$ from $\cA$, it is possible to introduce the scalar
product $\sigma_A : \cA \times \cA \longrightarrow \mathbb{C}$
by the rule
\be
\label{32}
\sgm_A : \; \left ( \hat A_1 , \; \hat A_2 \right ) = 
{\rm Tr}_{A} \hat A_1^+ \hat A_2   .
\ee
This scalar product generates the Hilbert-Schmidt norm
$|| \hat A || \equiv \sqrt{ \left ( \hat A_1 , \; \hat A_2 \right )}$.
The triple of the algebra of observables $\mathcal{A}$, acting on the 
Hilbert space $\mathcal{H}_A$, and the above scalar product $\sigma_A$
compose a Hilbert-Schmidt space
\be
\label{33}
\widetilde\cA \equiv \{ \cA, \; \cH_A , \; \sgm_A \} \;   .
\ee  

Let us introduce a composite Hilbert-Schmidt space by the tensor-product space
\be
\label{34}
 \widetilde\cA \bigotimes \widetilde\cB = \{ \cA, \; \cH_A , \; \sgm_A \}
\bigotimes \{ \cB, \; \cH_B , \; \sgm_B \}  \; .
\ee
An operator $\hat{C}$ in space (\ref{34}) is called separable if and only if
\be
\label{35}
  \hat C = \sum_i \hat A_i \bigotimes  \hat B_i \qquad 
( \hat A_i \in \widetilde\cA , \; \hat B_i \in \widetilde\cB ) \; ,
\ee
while it is entangled if and only if it cannot be represented in the 
separable form:
\be
\label{36}
  \hat C \neq \sum_i \hat A_i \bigotimes  \hat B_i \qquad 
( \hat A_i \in \widetilde\cA , \; \hat B_i \in \widetilde\cB ) \;  .
\ee

Prospects, being composite events, are represented, in view of 
Eq. (\ref{28}), by composite event operators. The structure of the prospect 
operators depends on how a composite Hilbert-Schmidt space is defined.
Generally, the prospect operators can be separable or entangled. Then the 
related prospects can also be termed separable or entangled. 

It is easy to give an example of a separable prospect. Let the algebras
$\mathcal{A}$ and $\mathcal{B}$ be composed of the corresponding projectors
$\hat{P}_n$ and $\hat{P}_\alpha$. The prospect $A_n \bigotimes B_\alpha$,
is represented by the relation
\be
\label{37}
 A_n \bigotimes B_\al \ra \hat P \left ( A_n \bigotimes B_\al \right ) =
\hat P_n \bigotimes  \hat P_\al \;  .
\ee
Here the prospect operator is clearly separable. Hence the prospect 
$A_n \bigotimes B_\alpha$ is called separable. Its probability is
\be
\label{38}
p \left ( A_n \bigotimes B_\al \right ) =  
{\rm Tr}_{AB} \hat\rho_{AB} \hat P_n \bigotimes \hat P_\al =
\lgl n \al | \hat\rho_{AB} | n \al \rgl \;   .
\ee

In contrast, entangled prospects appear when measurements or decision making 
are accomplished under uncertainty.

\subsection{Measurements and decisions under uncertainty}

An inconclusive event is a set $B = \{B_\alpha: \alpha =1,2,\ldots\}$ that 
is represented by a vector $|B \rangle$ of a Hilbert space, such that
\be
\label{39}
 B \ra | B \rgl = \sum_\al b_\al | \al \rgl \;  ,
\ee
with the event operator
\be
\label{40}
 \hat P_B = | B \rgl \lgl B | = 
\sum_{\al\bt} b_\al b^*_\bt  | \al \rgl \lgl \bt | \; .
\ee
In quantum measurements, an inconclusive event implies that after a measurement 
there is no a single measured value, but the result is a set of possible data 
$B_\alpha$ weighted with $|b_\alpha|^2$. In that sense, it is not a certain 
operationally testable event. In decision making, an inconclusive decision 
means that an exact decision is not yet actually taken, but it rather 
describes the process of deliberation between several possibilities, in that 
sense being an incomplete decision.  
 
Let us emphasize that an inconclusive event is not a union. Because an 
inconclusive event is represented as
\be
\label{41}
  B \ra \hat P_B = \sum_\al | b_\al|^2 \hat P_\al +
\sum_{\al\neq \bt} b_\al b^*_\bt  | \al \rgl \lgl \bt | \; ,
\ee
while a union is represented by the relation
\be
\label{42}
 \cup_\al B_\al \ra \sum_\al \hat P_\al \;  .
\ee
Therefore, the corresponding event operators are very different.   

One may say that an inconclusive event, being not uniquely operationally 
testable, cannot be the final stage of a measurement or decision making. 
But inconclusive events can occur, and often do exist, at intermediate 
stages of measurements and decisions. Actually, this is a typical situation 
for decisions under uncertainty. There are many such illustrations in the 
processes of physical measurement \cite{YS_24,YYS_34,YYS_35}
as well as in decision making \cite{YS_8,YS_16,YS_30,YS_31}. 

A typical prospect, describing a measurement or decision under uncertainty,
has the form
\be
\label{43}
  \pi_n = A_n \bigotimes B \; ,
\ee
where the final event $A_n$ is operationally testable, and $B = \{B_\alpha\}$
is an intermediate inconclusive event. This prospect is represented by the 
prospect state according to the relation
\be
\label{44}
\pi_n \ra | \pi_n \rgl = | n \rgl \bigotimes | B \rgl 
\ee
and induces the related prospect operator,
\be
\label{45}
  \pi_n \ra \hat P(\pi_n) = | \pi_n \rgl \lgl \pi_n | =
\hat P_n \bigotimes \hat P_B \; .
\ee
The explicit form of the latter is
\be
\label{46}
 \hat P(\pi_n) =  \sum_{\al \bt} b_\al b^*_\bt \hat P_n \bigotimes
| \al \rgl \lgl \bt | \;  .
\ee

The prospect states $|\pi_n \rangle$ are not necessarily orthonormalized. 
Therefore a prospect operator, generally, is not idempotent, since
\be
\label{47}
 \hat P^2(\pi_n) = \lgl \pi_n | \pi_n \rgl \; \hat P(\pi_n) \;  ,   
\ee
hence, it is not a projector. But the resolution of unity is required:
\be
\label{48}
 \sum_n \hat P(\pi_n) = \hat 1 \;  ,
\ee
where $\hat{1}$ is a unity operator in space (\ref{27}). The family of the 
prospect operators $\{\hat{P}(\pi_n)\}$ forms a positive operator-valued 
measure \cite{YS_24,YS_36}. 

The projectors $\hat{P}_n$ and $\hat P_\alpha$ represent operationally 
testable events. Because of this, the algebras of observables are defined
as the collections of these projectors. Thus, we have two algebras of 
observables
\be
\label{49}
\cA = \{ \hat P_n \} \; , \qquad \cB = \{ \hat P_\al \}
\ee
acting on the Hilbert spaces $\mathcal{H}_A$ and $\mathcal{H}_B$, 
respectively. With these algebras of observables in mind, we construct
the Hilbert-Schmidt space (\ref{34}). Then, analyzing the prospect
operator (\ref{46}), which can be written as
\be
\label{50}
 \hat P(\pi_n) = \sum_\al | b_\al|^2  \hat P_n \bigotimes \hat P_\al
+ \sum_{\al\neq \bt} b_\al b^*_\bt \hat P_n \bigotimes
| \al \rgl \lgl \bt | \; ,
\ee
we see that this operator is entangled, since, although the first term 
is separable, but the second term here is entangled. That is, prospect 
(\ref{43}) is also called entangled.

\section{Probability of uncertain prospects}

Suppose we consider several prospects forming a lattice
\be
\label{51}
\cL = \{ \pi_n : \; n = 1,2,\ldots, N \} \;   .
\ee
The probability of a prospect is given by the quantum form
\be
\label{52}
p(\pi_n) = {\rm Tr}_{AB} \hat\rho_{AB} \hat P(\pi_n) \;   .
\ee
By construction, the probability is non-negative and normalized,
\be
\label{53}
 \sum_n p(\pi_n) = 1 \;  , \qquad 0 \leq p(\pi_n) \leq 1 \;  ,
\ee
so that the family $\{p(\pi_n)\}$ is a probability measure. 

With the prospect operator (\ref{50}), it is straightforward to see that 
the prospect probability can be written as a sum of two terms,
\be
\label{54}
 p(\pi_n) = f(\pi_n) + q(\pi_n) \;  .
\ee
The first term $f(\pi_n)$ contains the diagonal part of Eq. (\ref{50}). It 
describes the objective utility of the prospect, because of which it is 
called the {\it utility factor}. The second term $q(\pi_n)$ is composed of 
the non-diagonal part of Eq. (\ref{50}) caused by the quantum nature of the 
probability. From the quantum-theory point of view, this term can be 
specified as an interference or coherence term. In decision theory, it 
characterizes subjective and subconscious feelings of the decision maker,
and can be named the {\it attraction factor} 
\cite{YS_8,YS_16,YS_30,YS_31}. 

It is worth stressing that form (\ref{54}) is not an assumption, but is the 
direct consequence of the definition for the prospect probability (\ref{52}), 
with the prospect operator (\ref{50}).  

By the quantum-classical correspondence principle, when the quantum term 
becomes zero, the quantum probability reduces to the classical probability,
so that
\be
\label{55}
 p(\pi_n) \ra f(\pi_n) \; , \qquad   q(\pi_n) \ra 0 \;  ,
\ee
with the normalization
\be
\label{56}
  \sum_n f(\pi_n) = 1 \;  , \qquad 0 \leq f(\pi_n) \leq 1 \;  .
\ee
In quantum theory, this is called decoherence.    

The attraction factor, by its construction, enjoys the following properties 
\cite{YS_8,YS_14,YS_16,YS_30,YS_31,YS_37}. It
lies in the range
\be
\label{57}
  -1 \leq q(\pi_n) \leq 1 \; 
\ee
and satisfies the {\it alternation law}
\be
\label{58}
 \sum_n q(\pi_n) = 0 \;  .
\ee
This law follows immediately from the form of probability (\ref{54}), 
under the normalization equations (\ref{53}) and (\ref{56}).

For a large class of distributions, there exists the {\it quarter law}
\be
\label{59}
 \frac{1}{N} \sum_{n=1}^N | q(\pi_n) | = \frac{1}{4} \;  .
\ee
The latter allows us to use as a non-informative prior the value 
$|q(\pi_n)| \approx 0.25$, which makes it possible to give quantitative 
predictions. 

Employing the definition of the conditional probability
\be
\label{60}
 p(A_n | B ) = \frac{p(A_n\bigotimes B)}{p(B)} \;  ,
\ee
for a prospect with an uncertain event $B$, we have
\be
\label{61}
p(A_n | B ) = 
\frac{\sum_\al |b_\al|^2 p(A_n\bigotimes B_\al)+q(\pi_n)}
{\sum_\al |b_\al|^2 p(B_\al)+q(B)}\;  .
\ee

The use of quantum probabilities is required when the quantum term $q(\pi_n)$ 
is not zero. As is clear from the above consideration, the necessary condition
for this is the occurrence of decisions under uncertainty. More precisely, 
the following theorem has been proved \cite{YS_24}: 

\vskip 2mm
{\parindent=0pt
{\bf Theorem.} {\it
For the quantum term $q(\pi_n)$ to be nonzero, it is 
necessary that the corresponding prospect $\pi_n$ be entangled and also 
the decision-maker state $\hat\rho$ be entangled.}     
}

\vskip 2mm
In the case of decisions under uncertainty, the prospect probability (\ref{54})
consists of two terms, utility factor and attraction factor. It is therefore
possible to classify the prospects from the given lattice (\ref{51}) in three
ways. A prospect $\pi_1$ is more useful than $\pi_2$, if and only if 
$f(\pi_1) > f(\pi_2)$. A prospect $\pi_1$ is more attractive than $\pi_2$, if
and only if $q(\pi_1) > q(\pi_2)$. And a prospect $\pi_1$ is preferable to 
$\pi_2$, if and only if $p(\pi_1) > p(\pi_2)$. In that way, a prospect can be 
more useful, but less attractive, as a result being less preferable, which 
explains all paradoxes in classical decision making 
\cite{YS_8,YS_16,YS_30,YS_31}.  

Let us stress that the principal difference of our approach in decision theory,
from all other models involving quantum techniques, is the possibility to not 
merely qualitatively interpret empirical results, but, moreover, to give their
quantitative description. As an example, let us briefly mention the prisoner 
dilemma game, where there are two prisoners who can either cooperate or 
defect (see details in Ref. \cite{Poundstone_38,Weibull_39,Kaminski_40}). Let 
$C_n$ denote cooperation, while $D_n$, defection. In our terminology, there 
are four separable prospects: $C_1  \bigotimes C_2$, $C_1  \bigotimes D_2$, 
$D_1  \bigotimes C_2$, and $D_1  \bigotimes D_2$. And the aim is to study the 
entangled uncertain prospects 
$$
 \pi_1 = C_1 \bigotimes \{ C_2 , \; D_2\} \; , \qquad
 \pi_2 = D_1 \bigotimes \{ C_2 , \; D_2\} \; ,
$$
corresponding to the choice between cooperation and defection for one of the 
prisoners, without knowing the decision of the other one. Empirical results 
of experiments, accomplished by Tversky and Shafir \cite{Tversky_41}, yield 
$p(\pi_1) = 0.37$ and $p(\pi_2) = 0.63$. In our approach, using the prior 
attraction factor $\pm 0.25$, we get $p(\pi_1) = 0.35$ and $p(\pi_2) = 0.65$, 
which, with the given experimental accuracy, coincides with the empirical data. 
A detailed description of this example can be found in Refs. \cite{YS_24,YS_42}.     

The prospect probabilities depend on the amount of available information. This
happens because the decision-maker strategic state depends on this information.
Let the information measure be denoted as $\mu$. The decision-maker states with 
this information and without it are respectively $\hat\rho(\mu)$ and
$\hat\rho(0)$. By the Kadison \cite{Kadison_43} theorem,
statistical operators, parameterized by a single parameter, are connected by 
means of a unitary operator $\hat{U}$ as
\be
\label{62}
 \hat\rho(\mu) = \hat U(\mu) \hat\rho(0) \hat U^+(\mu) \;  .
\ee
The prospect probability, with information $\mu$, is
\be
\label{63}
 p(\pi_n,\mu) = {\rm Tr} \hat\rho(\mu) \hat P(\pi_n) \;  .
\ee
Following the above consideration, we find that this probability is generally
the sum of two terms:
\be
\label{64}
  p(\pi_n,\mu) = f(\pi_n) + q(\pi_n,\mu) \; .
\ee
The first term, that is, the utility factor characterizes the prospect utility, 
and is not influenced by additional information, provided the utility is 
objectively defined. But the attraction factor, which is subjective, does 
depend on the available information. Employing the techniques used for 
treating the evolution of quantum systems \cite{Yukalov_44,Yukalov_50}, it 
is possible to show \cite{YS_37} that the attraction factor decreases with 
the received additional information approximately as 
\be
\label{65}
  q(\pi_n,\mu) \approx  q(\pi_n,0) e^{-\mu/\mu_c} \;  ,
\ee
where $\mu_c$ is the critical amount of information, after which the quantum
term strongly decays. 

The dependence of the attraction factor on the given information can explain
the effect of {\it preference reversal}. This effect was noticed by Tversky 
and Thaler \cite{Tversky_45}, who illustrated it by the following example. 
Imagine that people are asked to decide, under given conditions, between two 
programs, say $A$ and $B$. It may happen that they chose $B$ because it looks 
more useful. Then additional information is provided characterizing the cost 
of these programs. After getting this additional information, people choose 
$A$ instead of $B$, thus, demonstrating preference reversal. This effect is 
closely related to the planning paradox \cite{YS_16}. More detailed 
investigation of the preference reversal will be presented in a separate paper.

\section{Conclusion}

We have demonstrated the main mathematical points of a theory treating on the
same grounds both quantum measurements as well as quantum decision making.
The quantum joint and conditional probabilities have been introduced, being 
valid for arbitrary events, elementary as well as composite, operationally 
testable, as well as inconclusive, for commutative observables, as well as for 
non-commuting observables. The necessity of treating decision makers as members
of a society was emphasized. A pivotal point of the approach is the validity
of the quantum-classical correspondence principle that provides a criterion 
for constructing a correct and self-consistent theory. The necessary conditions
requiring the use of the quantum approach have been formulated. It was shown 
how additional information influences decision making. The developed Quantum 
Decision Theory does not meet paradoxes typical of classical decision making
and, moreover, makes it possible to give quantitative predictions. 
   
\section*{Acknowledgement}

One of the authors (V.I.Y.) is grateful to E.P. Yukalova for discussions.       

{\parindent=0pt 
\vskip 5mm
{\bf Funding}: Financial support from the Swiss National Foundation is appreciated.
}

\end{document}